\documentclass[aps,prl,preprint,byrevtex,groupedaddress,superscriptaddress,showkeys,titlepage,floatfix,balancelastpage,a4paper,10pt]{revtex4-1}

\usepackage{amsfonts,amssymb,amsmath,latexsym,graphicx}
\usepackage[english]{babel}
\usepackage{color}

\usepackage{url}

\def\beq {\begin{equation}}
\def\eeq {\end{equation}}

\newcommand{\w}{\omega}

\newcommand{\vq}{\mathbf{q}}

\begin{document}

% for line numbering, uncomment these four lines
% \linenumbers
% \setlength\linenumbersep{5pt}
% \renewcommand\linenumberfont{\normalfont\bfseries\tiny\textcolor{blue}}
% \modulolinenumbers[5]

\title{Revisiting the origin of satellites in core level photoemission of transparent conducting oxides: the case of $n$-doped SnO$_2$}

\author{Francesco Borgatti}
\affiliation{Consiglio Nazionale delle Ricerche -- Istituto per lo Studio dei Materiali Nanostrutturati (CNR-ISMN), via P. Gobetti 101, I-40129 Bologna, Italy}
\email{francesco.borgatti@cnr.it}

\author{J.A. Berger}
\affiliation{Laboratoire de Chimie et Physique Quantiques, IRSAMC, CNRS, Universit\'e Toulouse III - Paul Sabatier, 31062 Toulouse, France}
\affiliation{European Theoretical Spectroscopy Facility (ETSF)}

\author{Denis C\'eolin}
\affiliation{Synchrotron SOLEIL, L'Orme des Merisiers, Saint-Aubin, BP 48, F-91192 Gif-sur-Yvette, France}

\author{Jianqiang Sky Zhou}
\affiliation{Laboratoire des Solides Irradi\'es, \'Ecole Polytechnique, CNRS, CEA,  Universit\'e Paris-Saclay, F-91128 Palaiseau, France}
\affiliation{European Theoretical Spectroscopy Facility (ETSF)}

\author{Joshua J. Kas}
\affiliation{Department of Physics, University of Washington, Seattle, Washington 98195-1560, USA}

\author{Matteo Guzzo}
\affiliation{Institut f\"ur Physik und IRIS Adlershof, Humboldt-Universit\"at zu Berlin, D-12489 Berlin, Germany}

\author{C.F. McConville}
\affiliation{The University of Warwick, Dept. Phys., Coventry CV4 7AL, W Midlands, England}

\author{Francesco Offi}
\affiliation{Dipartimento di Scienze, Universit\`{a} di Roma Tre, 00146 Rome, Italy}

\author{Giancarlo Panaccione}
\affiliation{Istituto Officina dei Materiali CNR, Laboratorio TASC, S.S. 14 Km 163.5, AREA Science Park, 34149 Basovizza, Trieste, Italy}

\author{Anna Regoutz}
\affiliation{Inorganic Chemistry Laboratory, Department of Chemistry, University of Oxford, South Parks Road, Oxford OX1 3QR, United Kingdom}

\author{David J. Payne}
\affiliation{Department of Materials, Imperial College London, Exhibition Road, London SW7 2AZ, UK}

\author{Jean-Pascal Rueff}
\affiliation{Synchrotron SOLEIL, L'Orme des Merisiers, Saint-Aubin, BP 48, F-91192 Gif-sur-Yvette, France}
\affiliation{Sorbonne Universit\'es, UPMC Universit\'e Paris 06, CNRS, Laboratoire de Chimie Physique-Mati\'ere et Rayonnement, 75005 Paris, France}

\author{Oliver Bierwagen}
\affiliation{Materials Department, University of California, Santa Barbara, California 93106, USA}
\altaffiliation[Present address: ]{Paul-Drude-Institut f\"{u}r Festk\"{o}rperelektronik, 10117 Berlin, Germany}

\author{Mark E. White}
\affiliation{Materials Department, University of California, Santa Barbara, California 93106, USA}

\author{James S. Speck}
\affiliation{Materials Department, University of California, Santa Barbara, California 93106, USA}

\author{Matteo Gatti}
\affiliation{Laboratoire des Solides Irradi\'es, \'Ecole Polytechnique, CNRS, CEA,  Universit\'e Paris-Saclay, F-91128 Palaiseau, France}
\affiliation{European Theoretical Spectroscopy Facility (ETSF)}
\affiliation{Synchrotron SOLEIL, L'Orme des Merisiers, Saint-Aubin, BP 48, F-91192 Gif-sur-Yvette, France}
\email{matteo.gatti@polytechnique.fr}

\author{Russell G. Egdell}
\affiliation{Inorganic Chemistry Laboratory, Department of Chemistry, University of Oxford, South Parks Road, Oxford OX1 3QR, United Kingdom}

\date{\today}

\begin{abstract}
The longstanding problem of interpretation of satellite structures in core level photoemission spectra of metallic systems with a low density of conduction electrons is addressed using the specific example of Sb-doped SnO$_2$. Comparison of {\it ab initio} many-body calculations with experimental hard X-ray photoemission spectra of the Sn 4$d$ states shows that strong satellites are produced by coupling of the Sn core hole to the plasma oscillations of the free electrons introduced by doping. Within the same theoretical framework, spectral changes of the valence band spectra are also related to dynamical screening effects. These results demonstrate that, for the interpretation of electron correlation features in the core level photoelectron spectra of such narrow-band materials, going beyond the homogeneous electron gas electron-plasmon coupling model is essential. 

\end{abstract}


\maketitle

%%%%%%%%%%%%%%%%%%%%%%%%%%%%%%%%%%%%%
%\section{INTRODUCTION}
%%%%%%%%%%%%%%%%%%%%%%%%%%%%%%%%%%%%%
Transparent conducting oxides (TCOs) such as ZnO, CdO, SnO$_{2}$ and In$_{2}$O$_{3}$, combine optical transparency in the visible region with high electrical conductivity achieved through \emph{n}-type doping. These features enable a large variety of device applications in optoelectronics and photovoltaics, including their use as transparent electrodes in flat panel displays, organic light-emitting diodes, and solar cells \cite{Fortunato2007,Ellmer2012,Gao2016}. TCOs behave as dilute-electron systems whose density of conduction electrons is much lower than for simple metals, thus offering also an ideal platform to investigate the effects of electronic correlation through the interpretation of the satellite structures occurring in the X-ray photoemission spectra (PES) of core level and valence states. While it was recognized early on that plasmon satellites invariably accompany the main core level peaks in the photoemission spectra of simple metals \cite{Steiner1978,Huefner2003}, the interpretation  of  satellite structures observed in the spectra of  TCOs and other ``narrow band'' metallic oxides has remained controversial despite the large number of studies \cite{Campagna1975,Chazalviel1977,Edwards1984,Beatham1980,Cox1986,Cox1983,Egdell1999,Christou2000,Mudd2014}. After the emission of the photoelectron, as a direct consequence of  many-body interactions, the remaining electronic system can be left in different final states giving rise to several lines in the  spectrum. Satellites in PES are hence a genuine fingerprint of electronic correlation. In simple metals they were successfully explained on the basis of the homogeneous electron gas (HEG) model as the additional excitation of multiple plasmons \cite{Lundqvist1967_1,*Lundqvist1967_2,Langreth1970,Hedin1970,Hedin1999}, i.e. quantized charge density oscillations resulting from the long-range nature of the Coulomb interaction. In dilute-electron systems, plasmon energies are typically around 1 eV, which is much smaller than in simple metals and comparable both with the intrinsic core linewidths and with the chemical shifts associated to changes of the oxidation state. Moreover, while in simple metals -- as predicted by the HEG model -- the overall lineshape involves multiple plasmon loss satellites, for the ``narrow band'' metals only a single satellite is observed. Therefore, alternative explanations have been put forward, either in terms of mixed valency in the initial state \cite{DeAngelis1976,*DeAngelis1978}, or on the basis of localisation of conduction band states by the Coulomb potential of the core hole, leading to the appearance of screened and unscreened final states \cite{Kotani1974,Campagna1975,Chazalviel1977,Wertheim1979,DeGroot2008}.

Solving the ambiguity of the core double-peak structure problem for these systems calls for a joint experimental and theoretical advanced approach. To this end, in the present work we combine hard x-ray photoelectron spectroscopy (HAXPES) with {\it ab initio} many-body theory to explain the origin of satellites in the core level spectra of the prototypical undoped and Sb-doped SnO$_{2}$ TCO. We show that the experimental results and the calculated spectral functions correlate unambiguously the changes of the valence band (VB) spectrum and the satellite of the core level Sn 4$d$ peak to the intrinsic plasma oscillations of the free electrons that are introduced by charge carrier doping. The adoption of HAXPES ensures about the sensitivity to the electronic structure of the bulk, as it has recently emerged that the intensity of the final-state satellite structure in highly correlated metallic oxides (for example cuprate superconductors or doped manganites) is highly dependent on the information depth of the experiment and in some cases is only evident in the HAXPES measurements \cite{Panaccione2006,Offi2007,Offi2008,Fujii2011,Pincelli2017}. The calculations combine the GW approximation (GWA)  \cite{Hedin1965} for the self-energy and the cumulant (C) expansion of the Green's function into the first-principle GW+C scheme \cite{Martin2016}. This approach has lately been employed in simple metals and semiconductors (including doping effects)  \cite{Aryasetiawan1996,Kheifets2003,Guzzo2011,Gatti2013,Lischner2013,Guzzo2014,Lischner2015,Caruso2015,Zhou2015,Caruso2016,Verdi2017}, giving results in very good agreement with experiment. The application of this {\it ab initio} theory, which does not rely on any model assumption \cite{Kotani1974,Langreth1970}, to spectral features of different binding energy (BE) for a typical TCO material thus opens the perspective towards the reliable interpretation of electron correlation features in the PES of a large variety of TCO compounds and other conductive oxide materials.

%%%%%%%%%%%%%%%%%%%%%%%%%%%%%%%%%%%%%
% FIG 1 - EXPT. RESULTS
%%%%%%%%%%%%%%%%%%%%%%%%%%%%%%%%%%%%%

\begin{figure}[t]
\centering
\includegraphics[width=8 cm]{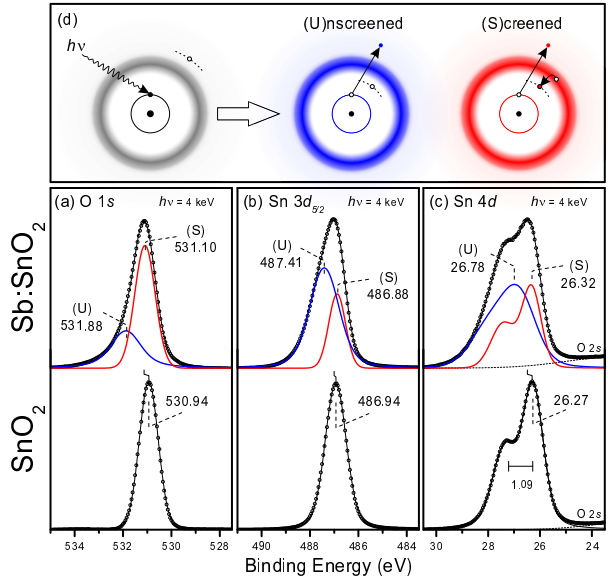}
\caption{(a-c) Core level HAXPES spectra of O 1$s$, Sn 3d$_{5/2}$ and Sn 4$d$ for nominally undoped SnO$_{2}$ (bottom) and Sb-doped SnO$_{2}$ (top), respectively. The background of the spectra has been subtracted using a Shirley profile. The peaks were fitted using Voigt profiles. (d) Scheme of the final-state configurations for the (S)creened and (U)nscreened channels accordingly to the Kotani-Toyazawa models \protect\cite{Kotani1974,DeGroot2008}.}
\label{Fig:FIG01}
\end{figure}

The HAXPES core level spectra in the O 1$s$, Sn 3d$_{5/2}$ and Sn 4$d$ region of nominally undoped SnO$_{2}$ and Sb-doped SnO$_{2}$ samples, grown by plasma assisted molecular beam epitaxy, are shown in Fig. \ref{Fig:FIG01}. The carrier concentration is 3.8x10$^{17}$ cm$^{-3}$ and 2.6x10$^{20}$ cm$^{-3}$, respectively \cite{White2008, White2009}. The spectra were obtained at the GALAXIES beamline of the SOLEIL synchrotron radiation source using quasi-monochromatic hard X-rays of about $h\nu = 4$ keV and normal emission detection geometry \cite{Rueff2015,Ceolin2013}. This condition provides an information depth of about 15 nm, large enough to ensure sensitivity to the electronic structure of the bulk and to minimize the contribution of the surface region. All the core level spectra for the doped sample exhibit striking broadening on the high BE side of the peaks,  with a small shift to higher binding energy of the peak maximum. Disentangling the spectral terms by Voigt functions in all cases shows a single peak for the undoped sample and two peaks separated by 0.71 eV (O 1$s$) and 0.53 eV (Sn 3$d_{5/2}$) for the doped one, while the shallow 4$d$ spectrum can be fitted by one or two Voigt pairs with a spin-orbit splitting energy of 1.09 eV and with an additional weak contribution from the O 2$s$ level, in agreement with previous XPS results for this and other TCOs \cite{Egdell1999,Mudd2014,Payne2007}. Further details of the fit analysis are reported in \cite{suppmat}.  In Fig. \ref{Fig:FIG01}(a-c) we have  named the spectral terms for Sb-doped SnO$_{2}$ according to the traditional description based on the  Kotani-Toyazawa models \cite{Kotani1974}, schematically shown in Fig. \ref{Fig:FIG01}(d), for which two different final states are accessible depending on whether a localised state pushed below the Fermi level remains empty, giving an ``unscreened'' state (U), or is filled by transfer of an electron from the conduction band to give a ``screened'' final state (S) \cite{Kotani1974,Campagna1975,Chazalviel1977,Wertheim1979}. In these models,  the screened final state gives rise to an asymmetric line to low BE side of the lifetime-broadened peak associated with the unscreened final state. We will now show how the {\it ab initio} theory that explicitly takes into account the dynamical screening of the photohole is able to explain these experimental findings without the need to resort to a  model approach.

%%%%%%%%%%%%%%%%%%%%%%%%%%%%
% FIG. 2 GW CALCULATION Sn 4d REGION
%%%%%%%%%%%%%%%%%%%%%%%%%%%%
The diagonal element of the spectral function for the $i$-th state  is the imaginary part of the one-particle Green's function: $ A_i(\w)= \pi^{-1} |\text{Im} G_i(\w)|$. 
In the cumulant expansion, merged with the GWA for the self-energy $\Sigma^{xc}$, the Green's function $G_i$ (for a hole state of energy smaller than the Fermi level $\mu$) is expressed as
%\beq
\begin{align}
 G_i(\w) =&  \frac{i}{2\pi} \int_{-\infty}^{0} dt e^{i(\w-\varepsilon_i)t}e^{C_i(t)} \label{eq:G} \\
 C_i(t) =&  \frac{1}{\pi}\int_{-\infty}^{\mu-\varepsilon_i}  d\w \frac{e^{-i{\w}t}-1}{\w^2} \text{Im} \Sigma_i^{xc}(\w+\varepsilon_i). \label{eq:cumulant}
\end{align}
%\eeq
Here, the quasiparticle (QP) energy $\varepsilon_i$ in the energy-self-consistent GW scheme is calculated as: $\varepsilon_i = \varepsilon^H_i + \text{Re}\Sigma^{xc}_i(\varepsilon_i)$, where $\varepsilon^H_i$ is the Hartree energy. In the GWA, the self-energy  $\Sigma^{xc}(\w)$ is the convolution of the Green's function $G(\w)$ and the dynamically screened Coulomb interaction $W(\w) = \epsilon^{-1}(\w)v_c$, with $v_c$ being the bare Coulomb interaction that is screened by the inverse dielectric function $\epsilon^{-1}(\w)$. Since in the GWA $\text{Im} \Sigma^{xc}(\w)$ is  proportional to the loss function $-\text{Im} \epsilon^{-1}(\w)$, the cumulant $C_i(t)$  describes the \emph{dynamical screening}  of the hole $\varepsilon_i$ through the coupling   with the bosonic charge excitations $\omega_s$, which are the plasmons and many-body interband transitions appearing as peaks in the loss function. Therefore, the first term in  Eq. \eqref{eq:G} yields the QP peak located at $\varepsilon_i$, while the exponential of $C_i(t)$ in \eqref{eq:G} generates satellite structures at multiples of the bosonic energies $\w_s$ away from the  QP peak. In the past, the $A_i(\w)$ of core levels have been extensively obtained using \emph{model} approaches \cite{Huefner2003,DeGroot2008}. In Ref. \cite{Langreth1970} Langreth demonstrated that for an isolated core level the cumulant expansion \eqref{eq:cumulant} yields the exact solution of the electron-boson model in the specific case of the HEG model. Kotani and Toyazawa \cite{Kotani1974} took into account the screening of the core hole by adopting a single-impurity Anderson model consisting of a localized state and conduction electrons. Instead, in the {\it ab initio} calculations for SnO$_2$ we can now make direct use of the band-structure information for the \emph{real} system. Doping is simulated by adding a HEG-type free-carrier contribution in the calculation of $\epsilon^{-1}(\w)$ \cite{suppmat}. To consider the doping dependence, the free-carrier concentration is similar to the experimental value as well as larger by one order of magnitude. 

\begin{figure}[t]
\centering
\includegraphics[width=5 cm]{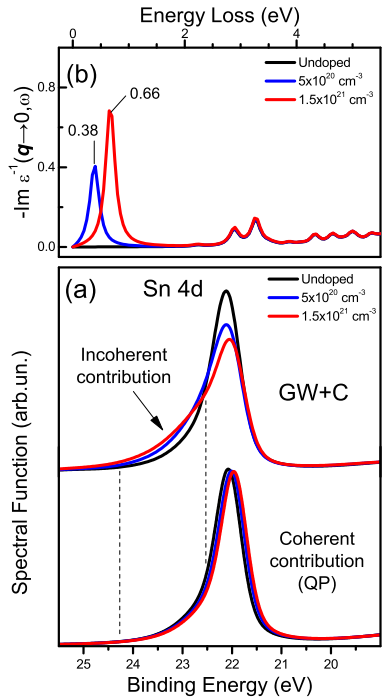}
\caption{(a) GW+C calculation of the spectral function for the Sn 4$d$ region of undoped (black) and doped (blue, red) compounds, the latter corresponding to two different carrier concentrations. The quasi-particle (QP) coherent contribution is shown in the bottom. (b) Loss function calculated for the same charge carrier concentrations. A plasmon peak arises as a consequence of doping.} 
\label{Fig:FIG02}
\end{figure}

The numerical results for the Sn 4$d$ core level, convoluted by a gaussian function with a FWHM of 0.47 eV corresponding to the experimental resolution of the HAXPES measurements, are shown in Fig. \ref{Fig:FIG02}(a) in which the QP contributions \cite{suppmat} have also been singled out from the total spectral functions calculated for Sn 4$d$ states \cite{foot}. The spin-orbit coupling is not included in the calculations, hence only a single peak is present. Noticeably, in the undoped case (black curves) the full spectral function coincides with the QP-only part: all the structures in the spectrum  are hence exclusively due to QP excitations. Upon doping the QP contributions remain the same and only a small shift of about 0.05 eV is found. Doping mostly affects the incoherent part of the spectral function: through a change of the dynamical screening $\epsilon^{-1}(\w)$, it induces a modification of electronic correlations \emph{beyond} the QP picture. Within the GW+C framework this correlation effect can be understood in terms of coupling of core hole and neutral excitations. In order to trace its origin, we have hence also calculated the loss function $-\textrm{Im} \epsilon^{-1}(\vq,\w)$ [see Fig.~\ref{Fig:FIG02}(b) for the $\vq \rightarrow 0$ limit]. For a finite carrier density we find a new plasmon peak in the loss function \cite{suppmat} at an energy corresponding to the separation between the QP and the new incoherent structures in the spectral function. We can safely conclude that the broadening of the Sn 4$d$ spectrum for the doped SnO$_{2}$ is the fingerprint of the correlation effect coupling the 4$d$ photoelectrons with the collective oscillation of the free carriers introduced in the material by doping. This mechanism should also be effective for the other core level spectra in Fig. \ref{Fig:FIG01}, although they cannot be simulated with our pseudopotential approach \cite{suppmat}.

In Fig. \ref{Fig:FIG02} we have also investigated what happens when the doping level is changed. According to the expectations based on the weak-coupling limit of the  HEG model \cite{Langreth1970}, the intrinsic plasmon satellite intensity for an isolated core level should increase as the conduction electron density $n_{c}$ decreases \cite{Hedin1999,Egdell2003}. At variance with these expectations, the intensity of the satellite decreases and merges with the QP broad structure making it hardly discernible. Moreover, in the {\it ab initio} calculations at most only one satellite is clearly visible in the spectra in agreement with HAXPES results, rather than a series of multiple plasmon satellites as predicted by the HEG model. These results therefore show that, in order to capture the satellite structures for narrow-band materials, it is essential to go beyond the HEG  electron-plasmon coupling model  Hamiltonian  \cite{Lundqvist1967_1,Lundqvist1967_2,Langreth1970} and to perform {\it ab initio} calculations that are materials specific \cite{suppmat}.

%%%%%%%%%%%%%%%%%%%%%%%%%%%%
% FIG. 3 Sn 4d SPECTRA THEORY VS. EXPT.
%%%%%%%%%%%%%%%%%%%%%%%%%%%%

\begin{figure}[t]
\centering
\includegraphics[width=5 cm]{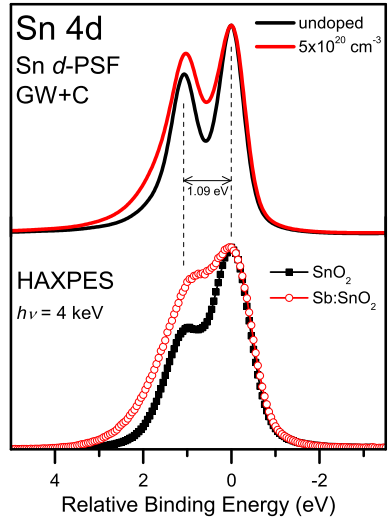}
\caption{Comparison of HAXPES Sn 4$d$ spectra (bottom) and $d$-orbital projection of the spectral functions (top) for undoped and doped SnO$_{2}$. The Shirley background and the small O 2$s$ term on the high BE side have been removed. All curves are aligned to the position of the experimental Sn 4$d_{5/2}$ term.}
\label{Fig:FIG03}
\end{figure}

The support for this conclusion comes through the close matching of theoretical and experimental results. To this aim, in Fig. \ref{Fig:FIG03} the Sn 4$d$ HAXPES spectra have been aligned to the maximum of the Sn 4$d_{5/2}$ peak to manifest more explicitly the broadening on the high BE side of the spectrum for doped SnO$_{2}$ after the subtraction of the background and of the O 2$s$ contribution close to the high BE side of the region. The spectra are compared to the $d$-orbital projection of the spectral functions for which the spin-orbit splitting has been phenomenologically included by adding the same curve shifted for the experimental spin-orbit value (1.09 eV) and scaled to achieve the statistical branching ratio of $d$-doublets. The calculations have also been gaussian-convoluted to account for the experimental resolution. Despite the implicit rough approximations of this treatment, the qualitative agreement with experiment is very good, especially considering that extrinsic effects, due to inelastic losses of the outgoing photoelectrons, and their interference with intrinsic contributions, are not included in the calculations \cite{Hedin1998,Huefner2003}.

%%%%%%%%%%%%%%%%%%%%%%%%%%%%
% FIG. 4 VB SPECTRA THEORY VS. EXPT.
%%%%%%%%%%%%%%%%%%%%%%%%%%%%
\begin{figure}[t]
\centering
\includegraphics[width=6 cm]{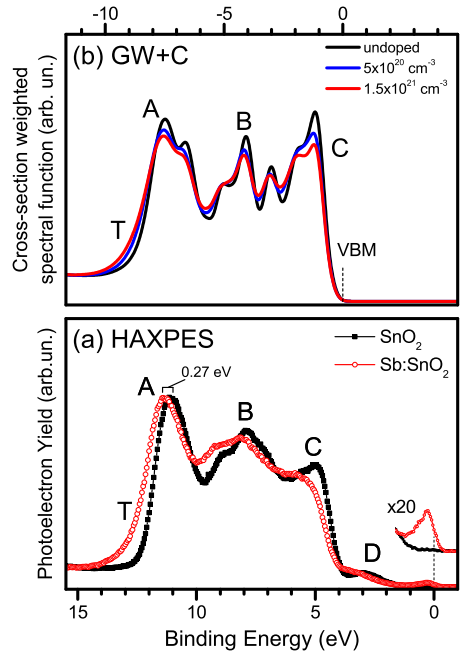}
\caption{(a) HAXPES VB spectra normalized to the maximum intensity of the stronger features. The spectral contribution of the conduction states for the doped sample has been enhanced by a factor 20. (b) GW+C spectral functions whose angular-projected contributions have been weighted by the respective photoionization cross sections. The energy scale of the calculations is referred to the Valence Band Maximum (VBM) to highlight the doping-induced tail feature (T) \cite{foot}.}
\label{Fig:FIG04}
\end{figure}

Further, in Fig.  \ref{Fig:FIG04} we consider the valence band (VB) spectra. For both the pure and the doped compounds, the HAXPES VB spectra display a three-peaked structure (A-C) in agreement with XPS data previously reported \cite{Egdell1999}. For the undoped sample, the onset of the valence band maximum (VBM) is placed through linear extrapolation at about 3.84 eV binding energy relative to the Fermi level E$_{F}$, slightly larger than the quoted bulk band gap of 3.62 eV \cite{Batzill2005}. On the low BE side, the spectra have a weak but well-defined feature (D) that tails into the bulk band gap, which was interpreted as a contribution from lone-pair Sn 5$s$-states occurring at the sample surface and, for the doped sample, concerns also Sb 5$s$-5$p$ states \cite{Cox1982}. Looking at the position of the stronger features, the spectrum of the Sb-doped SnO$_{2}$ is shifted to higher BE by 0.27 eV, and the shift of the conduction states below E$_{F}$ is clearly visible as well as the broadening of the tail (T) on the high BE side. The comparison of the VB states with the calculated spectral function, whose angular-projected contributions have been weighted by the correspondent values of the photoionization cross sections \cite{Scofield1973}  and gaussian-convoluted accordingly to the experimental resolution, shows that both strength and position of the spectral features are quite precisely reproduced by the calculations, suggesting also that most of the VB spectral weight for HAXPES measurements derives from the Sn contribution \cite{foot}. The calculations reproduce the broadening of all the peaks induced by the free-carrier doping. Notably, similarly to the experiment, the asymmetric tail at the bottom of the VB is clearly enhanced. This observation nicely confirms Hedin's predictions based on the HEG for valence electrons \cite{Hedin1980}. In the past, asymmetric tails have often been modeled for core levels using {\it ad hoc} Doniach-Sunjic or Mahan lineshapes  \cite{Doniach1970,Mahan1975,Huefner2003}. In the present case, instead, this result emerges consistently from the same {\it ab initio} framework that is used for \emph{both} the Sn 4$d$ states and the VB. As for the satellite of the Sn 4$d$ core line, this tail can be associated with dynamical screening effects requiring to go beyond a single-particle picture. This is a direct consequence of the frequency dependence of $\text{Im} \Sigma^{xc}(\w)$ \cite{Hedin1970,Hedin1980,Hedin1999}, which is related to the presence of low-energy ($\sim$ 1-2 eV) density fluctuations that couple with high-energy QP excitations in the photoemission spectra \cite{Gatti2013,Gatti2015}.

%%%%%%%%%%%%%%%%%%%%%%%%%%%%
%CONCLUSIONS
%%%%%%%%%%%%%%%%%%%%%%%%%%%%
In conclusion, we have elucidated the origin of the satellite structure observed in the Sn 4$d$ core level photoemission spectrum of Sb-doped SnO$_2$ by comparing experimental measurements to results obtained from {\it ab initio} many-body theory. We have demonstrated that such a satellite is produced by the coupling of the Sn 4$d$ core electrons to the plasma oscillation of the free electrons introduced in the material by doping. Moreover, within the same theoretical framework we were able to explain also the enhancement of the asymmetric tail in the valence band photoemission spectrum of doped SnO$_2$. These results demonstrate that, in order to capture the satellite structures for narrow-band materials and identify properly the underlying electronic structure excitations, it is essential to go beyond the HEG electron-plasmon coupling model and to perform material-specific {\it ab initio} calculations. In this perspective, the results for the Sb-doped SnO$_2$ TCO suggest that the GW+C theory can be a very promising approach for the interpretation of electron correlation features in the PES spectra of several conductive oxide materials.

\begin{acknowledgments}
This research was supported by a Marie Curie FP7 Integration Grant within the 7th European Union Framework Programme. Computational time was granted by GENCI (Project No. 544). Supporto for JJK was provided by DOE BES Grant No. DE-FG02-97ER45623.
\end{acknowledgments}

\bibliography{biblio}

\begin{thebibliography}{61}
\bibitem [{foo()}]{foot}%
  \BibitemOpen
  \href@noop {} {}\bibinfo {note} {Note that the calculated spectral functions
  have been aligned to their valence band maximum and not to the Fermi level,
  as in the experiment, since the work function cannot be determined in
  calculations for the bulk. Moreover, the binding energy for shallow Sn $4d$
  core levels could be improved by QP self-consistency and inclusion of vertex
  corrections \protect\cite{Schilfgaarde2006,Bruneval2014,Gruneis2014}, however
  this is not crucial for the present work.}\BibitemShut {Stop}%
\end{thebibliography}

\end{document}